\documentclass{iauc}
\usepackage{graphicx}

\title[Luminosity Function] 
{Environmental Dependencies in the Luminosity Function of Galaxies}

\author[R.B. Tully]   
{R. Brent Tully$^1$}

\affiliation{$^1$Institute for Astronomy, University of Hawaii, Honolulu,
HI 96822, USA \break 
email: tully@ifa.hawaii.edu}

\pubyear{2005}
\volume{198} 
\pagerange{000--000}
\date{?? and in revised form ??}
\jname{Near-Field Cosmology with Dwarf Elliptical Galaxies}
\editors{H. Jerjen and B. Binggeli, eds.}
\begin{document}

\maketitle

\begin{abstract}

The evidence is becoming strong that the luminosity function of galaxies
varies with environment.  Higher density, more dynamically evolved
regions appear to have more dwarfs per giant.  The situation is becoming
clearer as a result of wide field imaging surveys with the 
Canada-France-Hawaii and Subaru telescopes and spectroscopy of faint
dwarfs with the Keck telescope.  We report here on extensive observations
of the small but dense NGC~5846 Group.  The faint end of the luminosity
function rises relatively steeply in this case.


\end{abstract}

\firstsection 
\section{Introduction}

At the inception of this program, circa 1997, the following thoughts 
provided motivation.
The properties of the luminosity function of galaxies at the time were poorly 
constrained by observations at the faint end.
Most of the observational attention had been directed toward rich clusters
of galaxies because they could provide good statistics and contrast against
interlopers.
However rich clusters are relatively distant and at faint levels the 
luminosity function is only defined after statistical corrections for
contamination.
Experience in and near the Local Group teaches us that the vast majority of
dwarf galaxies are low surface brightness, a property that distinguishes
them from background objects.  If we confine a study to groups that are
clean from the perspective of confusion from nearby structures then we might
reliably identify group members without need for statistical corrections.
By looking at nearby groups, we can observe down to faint magnitudes, over
a variety of environments including some where the density is low.

We began with a study of the Ursa Major (UMa) Cluster because we were familiar
with the region from work on the bright members 
(\cite{tul96}; \cite{ver01}).  The UMa Cluster almost qualifies as Abell
richness class 0 but it is different from most clusters to receive attention
because it is irregular, low density, and spiral rich.  The cluster is at
roughly the same distance as the Virgo Cluster and subtends 100 sq. deg.
We sampled 15\% of the area with strips along the major and minor axes of
the elongated cluster with observations using the 12k imager at the 
Canada-France-Hawaii Telescope and the neutral hydrogen receiver at the 
Very Large Array.  We conducted a double blind search for dwarf galaxies
with Neil Trentham conducting the optical search and Marc Verheijen conducting
the HI search.

\begin{figure}
\begin{center}
\scalebox{.4}{
 \includegraphics{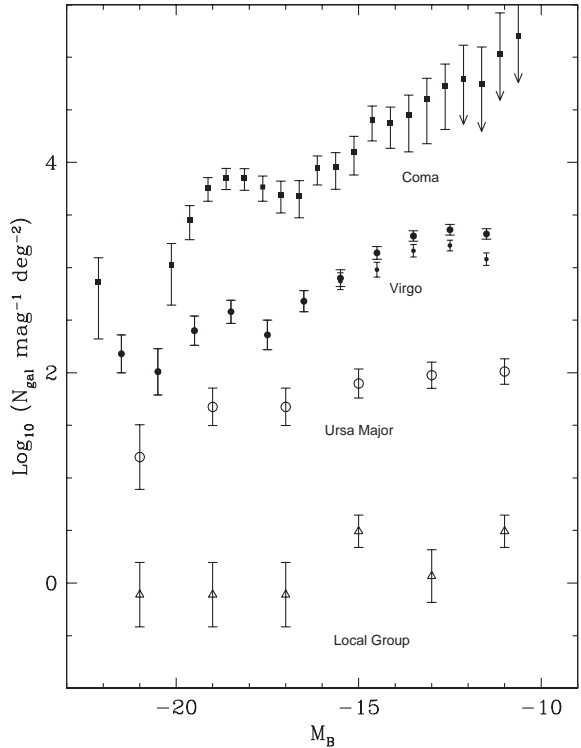}
 }
\end{center}
  \caption{
$B$-band luminosity functions for the Coma, Virgo, Ursa Major clusters and
the Local Group.  Vertical scales are shifted for clarity.  Downward arrows
in the case of the Coma Cluster reflect background contamination 
uncertainties.
}\label{fig:lf_4groups}
\end{figure}

The HI survey turned up all the previously known galaxies including a couple
that had excaped detection in HI but only uncovered 11 new objects.  The
HI mass flunction in UMa was demonstrated to be flat at low masses
(\cite{ver00}).  The optical survey turned up 3 times as many candidates,
still not a large number.  Probably many, if not most, of these candidates
would have been detected in HI if the sensitivity achieved with the
VLA could have been pushed a factor 10 below our $10^7 M_{\odot}$ limit.
In any event, the optical luminosity function in UMa is almost flat at the
faint end (see Figure~\ref{fig:lf_4groups} extracted from \cite{tre01}).

\section{Environmental Dependencies?}

The luminosity function in the UMa Cluster is flatter than others have found
in denser, more dynamically evolved places, reminiscent of the situation
found in the Local Group (\cite{kly99}; \cite{moo99}).  The luminosity
function in these low density environments is {\it much flatter} than the mass
function anticipated by $\Lambda CDM$ hierarchical clustering theory
(\cite{she99}).   The theory predicts that the mass spectrum is {\it steeper}
in lower density environments, not flatter (\cite{sig00}).  What could explain
an increased departure of the luminosity function from the mass function in
lower density environments?

\cite{tho96}, \cite{gne00} and others have discussed the way reionization
could have suppressed the accumulation of gas in small dark matter halos.
Hot intergalactic gas has too much energy to fall into a small dark matter
halo that collapses after the epoch of reionization.  By contrast, a small
halo that collapses before reionization will be accompanied by cold gas.
Early forming low mass halos would contain gas and form stars while late 
forming low mass halos would not contain gas, hence not be the sites of 
star formation.  This scenario could explain the trends seen in the properties
of the luminosity function because low mass halos collapse earlier in 
environments that become rich clusters than in low density environments
like those that become the Local Group or the Ursa Major Cluster 
(\cite{tul02}).  Semi-analytic models suggest that 
gas retention is strongly favored in
places that become dense, massive clusters because some halos form very early.
Even in these densest places, though, {\it most} 
halos will form after reionization; ie, gas will not be pulled in with the 
collapse of most halos.

\section{Pilot Program to Survey a Variety of Environments}

Our next step was to take a quick look at a wide variety of environments 
within the Local Supercluster.
Subaru SuprimeCam was used to image the inner parts of 5 additional groups,
which added to UMa gave us 6 samples (\cite{tre02}). Trends were seen in 
the suspected
sense: more dynamically evolved regions have higher dwarf/giant ratios.
The mean slope at the faint end of the luminosity function is characterized
by a \cite{sch76} slope parameter $\alpha = -1.19 \pm 0.03$.  However, 
not enough sky was surveyed in each group for sufficient statistics on
individual groups.

\begin{figure}
\begin{center}
\scalebox{.4}{

 \includegraphics{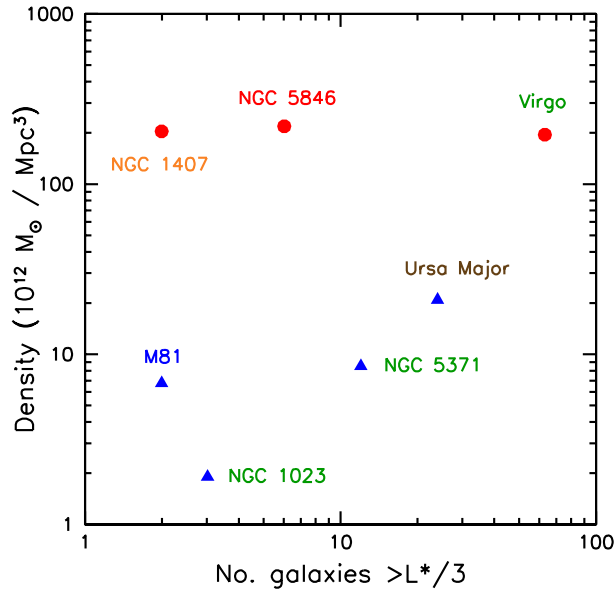}
 }
\end{center}
  \caption{
Seven cluster or group environments will be surveyed over the course of
this program.  The environments are characterized by a density parameter
(virial mass / volume) and a richness parameter (number of members more
luminous than $L^{\star}/3$).  Groups of predominantly early-type 
and late-type galaxies
are identified by circles and triangles, respectively.
The present paper summarizes results
obtained for the dense but relatively poor NGC~5846 Group.
}\label{fig:program}
\end{figure}

The problem required much more sky coverage.  Entire groups should be surveyed
and groups with very different properties should be considered.  We began
a program with the wide field imagers on the Canada-France-Hawaii Telescope,
culminating in the use of the 1 sq. deg. field MegaCam.  Fields were chosen
to maximize the range of group properties from numerically rich to poor
in bright galaxies 
and from dense and dynamically evolved to sparse and dynamically
young.  The seven groups identified in Figure~\ref{fig:program} have been 
or will be observed.  We report here on the 
case of the NGC~5846 Group.

\section{The Small, Dense NGC~5846 Group}

The number of luminous galaxies in the NGC~5846 Group is modest, with 
only 3 $L^{\star}$ galaxies, but the group is
dynamically evolved, with almost all members of early type.  It is a 
dumbbell system: members congregate around the two bright ellipticals
NGC~5813 and NGC~5846.  Extended X-ray emission is centered on these two
major galaxies.  The distance to the group is 26 Mpc.  The velocity
dispersion, from 86 redshifts, is 322 km/s, resulting in a virial mass
estimate of $8 \times 10^{13} M_{\odot}$ and mass to light ratio of 320
in solar units.   

The definition of the luminosity function to faint levels depends on the
ability to identify group members free of contamination.  We first select
candidates based on quantitative surface brightness criteria that would 
assure inclusion of dwarfs seen in the local neighborhood and 
redshift-confirmed members of the UMa and Virgo clusters.  These low 
surface brightness candidates were then evaluated on morphological grounds
and given qualitative group membership ratings (1) probable, (2) possible,
(3) plausible, and (4) implausible.

Our rating scheme was then tested in two ways.  The first and most rigorous
test was to
obtain new velocities.  The Sloan Digital Sky Survey (SDSS) now
provides redshifts for essentially all our galaxies with $R<17$ ($M_R < -15$).
We have obtained additional spectra  for galaxies up to 2 mag fainter with
Keck Telescope.  It was found that (a) all 24 galaxies with new velocities
that were rated 1 and 2 are
members, (b) 16 of 23 galaxies rated 3 are members, (c) only 1 galaxy rated
4 has a new velocity that shows it to be a member, and (d) 5 high surface 
brightness galaxies have turned out to
be members.  The second test involves intercomparisons of spatial 
correlations.  Galaxies rated 1 and 2 are as highly correlated with the 
group as galaxies confirmed as members on the basis of velocities, indicating
that essentially 100\% of these objects belong to the group.  By contrast,
the correlation with the galaxies rated 3 indicates that $50\% \pm 10\%$
of these candidates are in the group.  It is concluded that $255 \pm 15$
of our candidates are associated with the group.  Of the dwarfs, 80\% are
early types.

There turns out to be a clean separation between high and low surface
brightness objects in plots of surface brightness vs. luminosity.
There are 3 very compact ellipticals; one rather like M32.  It turns out
that these three unusual objects are the {\it nearest neighbors} in
projection to the
dominant NGC~5846 galaxy.  We suggest that these 3 galaxies have been tidally
truncated.

\begin{figure}
\begin{center}
\scalebox{.5}{
 \includegraphics{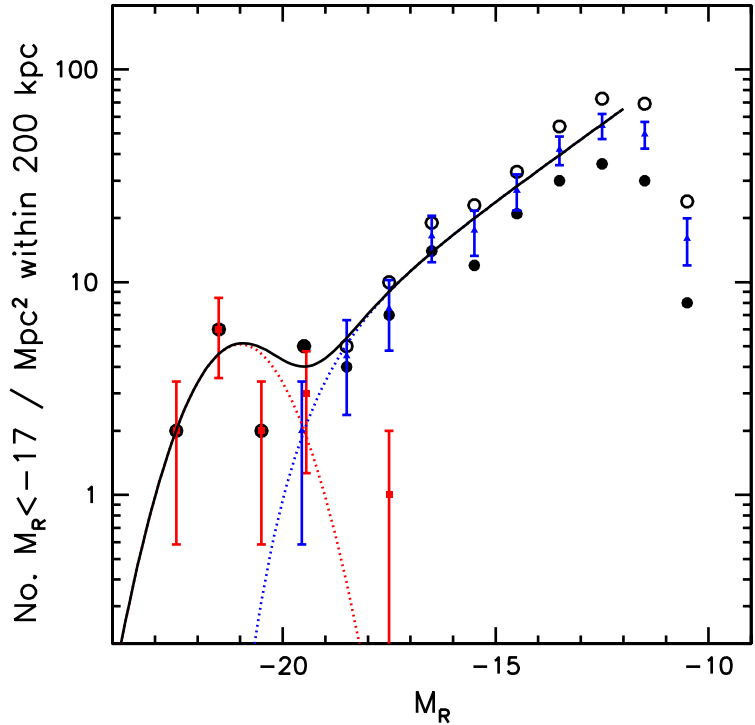}
 }
\end{center}
  \caption{
NGC~5846 Group luminosity function normalized by the surface density of
galaxies with $M_R < -17$ at 200~kpc radius.  Large filled circles: number
of galaxies per half-magnitude bin either confirmed as a group member with
a redshift or almost a certain member with a morphological rating 1 or 2.
Large open circles: same except also including morphological rating 3.
Small filled squares: high surface brightness only.  Small filled triangles:
low surface brightness only, splitting the difference between the values
represented by the filled and open large circles.  Dotted curves: separate
fits to the high and low surface brightness distributions.  Solid curve:
sum of the two component fits.  Parameters of the fit: $M_g = -21.0 \pm 0.3$,
$\sigma_g = 1.1 \pm 0.2$, $N_g = 5.1 \pm 1.2$, $M_d = -19.5 \pm 0.6$,
$\alpha_d = -1.36 \pm 0.04$, $N_d = 5.9 \pm 0.4$.
}\label{fig:lf}
\end{figure}

The identification of group members allows us to build the luminosity 
function seen in Figure~\ref{fig:lf}.  There is a saddle point in the 
observed distribution at 
$M_R \sim -20$, similar to a feature seen in other samples, and not something
that can be described by a Schechter function.  Following our procedure in
the Subaru pilot program (\cite{tre02}), we fit the luminous, high 
surface brightness galaxies with a Gaussian function ($M_g=-21.0$, 
$\sigma_g = 1.1^m$) and fit the faint,
low surface brightness component with a Schechter function 
($M_d^{\star} = -19.5$, $\alpha_d = -1.36$).  The luminosity function is
well defined to a completion limit of $M_R = -12$.

\begin{figure}
\begin{center}
\scalebox{.4}{
 \includegraphics{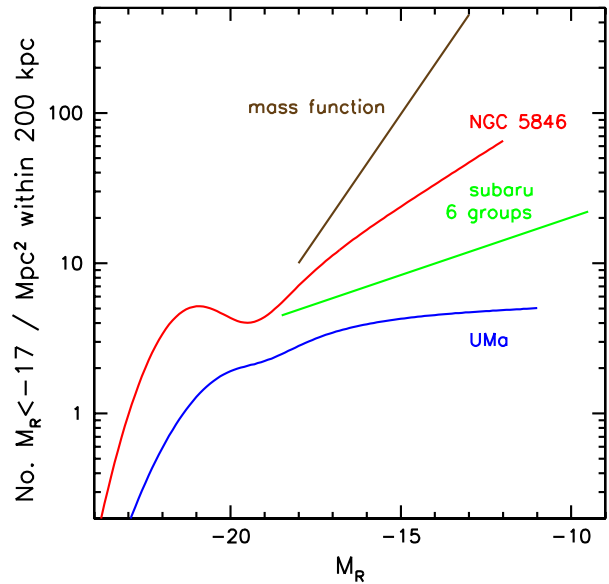}
 }
\end{center}
  \caption{
Schematic comparison of the NGC~5846 Group luminosity
function with other cases.  The curve representing the NGC~5846 Group
is the solid curve in Fig.~\ref{fig:lf}.  The curves labeled
`subaru 6 groups' and `UMa' are taken from \cite{tre02}.  The 6 groups
curve represents an average found for the dwarf regime in 6 groups
discussed by \cite{tre02}.  The density normalizations are carried out in a
consistent way in these separate cases.
The steeper line labeled `mass function' indicated the low mass slope
of a modified Press-Schechter distribution
with arbitrary vertical scaling.
}\label{fig:lf_compare}
\end{figure}

The luminosity function for the NGC~5846 Group is compared with what was 
found
from other environments in Figure~\ref{fig:lf_compare}.
The faint end slope of $\alpha = -1.36 \pm 0.04$ is significantly steeper
than the mean slope for the 5 Subaru groups of $\alpha = -1.19 \pm 0.03$,
and steeper still than the slope $\alpha \sim -1.0$ found for the UMa
sample.  However it is shallower than the Press-Schechter mass function
slope.  The NGC~5846 Group environment is dense and dynamically evolved, 
even though the number of bright galaxies is modest and the total mass 
is not extreme.  Our information is still limited but we are seeing a
pattern that appears consistent with the reionization model (\cite{tul02}).
Details of the NGC~5846 study are presented by \cite{mah05}.

\begin{acknowledgments}
My collaborators in this work are Neil Trentham, Andisheh Mahdavi, 
Rachel Somerville, and Marc Verheijen.  Support is provided by the US
National Science Foundation award AST-03-07706.
\end{acknowledgments}

\end{document}